\documentclass[floatfix,aps,prl,twocolumn,superscriptaddress,showpacs]{revtex4-1}
\usepackage[normalem]{ulem}
\usepackage{float}
\usepackage[usenames]{color}
\usepackage{graphicx}
\usepackage{ulem}
\usepackage{subfigure}
\usepackage{epstopdf}
\usepackage{amsfonts}
\usepackage{amsmath}

\begin{document}

\title{Nonequilibrium dynamics of probe filaments in actin-myosin networks}

\author{J. Gladrow}
\affiliation{Third Institute of Physics, Georg August University, 37077 G\"{o}ttingen, Germany }
\affiliation{Cavendish Laboratory, University of Cambridge, Cambridge CB3 0HE, United Kingdom}
\author{C.P. Broedersz}
\email{C.broedersz@lmu.de}
\affiliation{Arnold-Sommerfeld-Center for Theoretical Physics and Center for
  NanoScience, Ludwig-Maximilians-Universit\"at M\"unchen,
   D-80333 M\"unchen, Germany.}
\affiliation{Kavli Institute for Theoretical Physics, University of California, Santa Barbara, California 93106, USA}

\author{C.F. Schmidt}
\email{christoph.schmidt@phys.uni-goettingen.de}
\affiliation{Third Institute of Physics, Georg August University, 37077 G\"{o}ttingen, Germany }
\affiliation{Kavli Institute for Theoretical Physics, University of California, Santa Barbara, California 93106, USA}

\date{\today}

\begin{abstract}
Active dynamic processes of cells are largely driven by the cytoskeleton, a complex and adaptable semiflexible polymer network, “motorized” by mechanoenzymes. Small dimensions, confined geometries and hierarchical structures make it challenging to probe dynamics and mechanical response of such networks. Embedded semiflexible probe polymers can serve as non-perturbing multi-scale probes to detect force distributions in active polymer networks. We show here that motor-induced forces transmitted to the probe polymers are reflected in non-equilibrium bending dynamics, which we analyze in terms of spatial eigenmodes of an elastic beam. We demonstrate how these active forces induce correlations among these mode amplitudes, which furthermore break time-reversal symmetry. This leads to a breaking of detailed balance in this mode space. We derive analytical predictions for the magnitude of resulting probability currents in mode space in the white-noise limit of motor activity. We relate the structure of these currents to the spatial profile of motor-induced forces along the probe polymers and provide a general relation for observable currents on two-dimensional hyperplanes.   
\end{abstract}

\maketitle

\noindent

The emergent field of active matter aims to develop systematic descriptions of stochastic out-of-equilibrium phenomena in energy-dissipating soft matter systems~\cite{MacKintosh2010,Marchetti2013, Prost2015}. A prominent motivation for such studies are the dynamics observed in living cells and tissues. \emph{In vitro} model systems based on one of the main ingredients of the cellular cytoskeleton, filamentous actin, have been playing a prominent role in pioneering experimental studies of active matter. Building on earlier microrheology experiments on equilibrium actin networks~\cite{Ziemann1994, Schnurr1997, Gittes1997, Hinner1998, Gardel2003, Koenderink2006, MacKintosh2010}, similar experiments in reconstituted actin networks including myosin motor proteins have revealed that motor activity can drastically alter the mechanical response of actin networks~\cite{Lau2003, Liu2006, Koenderink2009, Broedersz2011, Ronceray2016} and significantly enhance fluctuations~\cite{Brangwynne2008, Mizuno2007, Mizuno2009}. Actin-myosin model systems have also been used to study structural self-organization and pattern formation~\cite{SoareseSilva2011, Murrell2012, Alvarado2013, Lenz2014, Weber2010, Schaller2011}. 

The standard approaches to characterizing cellular dynamics involve direct microscopic imaging of the motion of whole cells or identifiable features within cells, or the monitoring of displacements of probes inserted into cells or attached to the outside of cells. Small probes are often tracked via fluorescence microscopy. Micron-sized colloidal beads attached to cells or injected into cells have been used to track fluctuations~\cite{Fabry2001, Hoffman05072006, Deng2006, Mizuno2009, Guo2014}. However, these probes clearly lack spatial resolution and often cannot enter confined geometries such as the actin cortex of non-adherent cells. An interesting alternative is the direct fluorescent labeling of parts of the cytoskeleton itself, for example microtubules~\cite{Brangwynne2008} or the insertion of high-aspect-ratio filamentous probes, such as fluorescent single-walled carbon nanotubes~\cite{Fakhri2010, Fakhri2014, Tan2016} that can penetrate into tight spaces in the cell. 
\begin{figure}
  \begin{center}
    \includegraphics[width=\columnwidth]{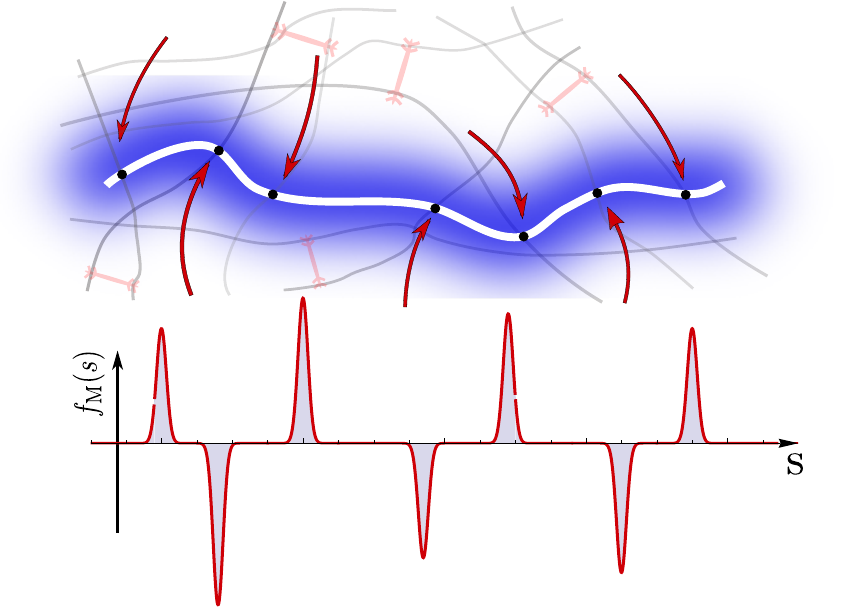}
    \vspace{-0.1in}
    \caption{(color online)  Sketch of the scenario of probe filament in motor activated network. A probe filament, e.g. a carbon nanotube (blue), is introduced into a crosslinked network of actin filaments (grey), with myosin motors (red). Motor action (red arrows) results in external forces $f_\text{M}$ impinging on the probe along its contour $s$. Both, myosin and actin are drawn opaque as they would not be visible in fluorescence microscopy experiments. }
\label{fig:networks}
  \end{center}
\end{figure}
Once the stochastic dynamics are tracked in an active material, it can be challenging to distinguish equilibrium from non-equilibrium stochastic motions and to quantify the extent of activity. Colloids in active networks typically exhibit non-Gaussian displacement distributions~\cite{Toyota2011,Ben-Isaac2015, Turlier2016}, but that feature by itself does not prove non-equilibrium. Scaling regimes of non-equilibrium filament fluctuations have been theoretically investigated and are expected to deviate from equilibrium predictions~\cite{Everaers1999}. If fluctuations can be compared to material response properties, the fluctuation-dissipation theorem can be applied to quantify activity ~\cite{Hoffman05072006, Mizuno2007, Mizuno2008}. Recently, a non-invasive method was introduced to discern active fluctuations based on the violation of detailed balance ~\cite{Battle2016, Gladrow2016}. This method can be applied on standard microscopic imaging data, and is based on analyzing probability flux patterns in phase spaces constructed from two or more degrees of freedom  of the system, which are either directly coupled or share their driving forces. 

In contrast to point-like probe particles or spherical beads, extended filaments offer an easily accessible spectrum of simultaneously observable variables: their bending modes~\cite{aragon1985dynamics, Gittes1993, Brangwynne2007}.
We can decompose the instantaneous filament conformation into a sum of dynamic bending eigenmodes with the following properties: (i) Particular spatial modes act as reporters of dynamics at their respective characteristic length and time scale; since viscous relaxation times are length dependent, modes also have characteristic relaxational timescales. (ii) In equilibrium, dynamic eigenmodes are statistically independent and each mode amplitude individually fulfils a fluctuation-dissipation theorem. 
In a motor-activated network, by contrast, a filament will receive random kicks from the network generated by myosin motors and transmitted by the network. Such motor activity results in enhanced mode amplitude fluctuations and may, as we will show, abrogate the independence of the eigenmodes. In a preceding letter~\cite{Gladrow2016}, we have shown that resultant mode cross-correlations are accompanied by circular patterns of probability fluxes in mode space, which indicate a breakdown of detailed balance. In this way, the analysis of bending eigenmodes of embedded probe filaments can be used to indicate non-equilibrium dynamics in an active polymer network.

In this paper, we develop the analysis of filament motion in an active viscoelastic medium in more detail and specify how the spatial structure of actively induced network-probe interactions, as sketched in Fig.~\ref{fig:networks}, translates into mode cross-correlations. We discuss the breaking of Onsager's time reversal symmetry in this system and show how cross-correlations can provide guidance to identify motor-induced dynamics. Furthermore, we also discuss an additional non-equilibrium marker in the white-noise limit, the frequency associated with the circulatory probability currents that arise in non-equilibrium steady-states.


\section{The filament model}
\label{sec:model}
\begin{figure}
  \centering
  \includegraphics[width=\columnwidth]{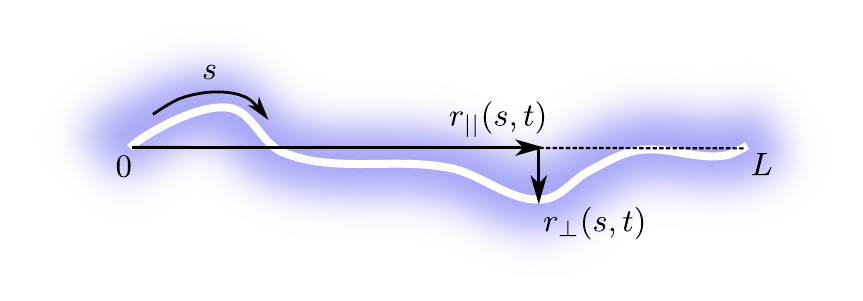}
  \caption{Parametrization of the probe filament shape.}
  \label{fig:parametrizationOfProbe}
\end{figure}

We consider the non-equilibrium dynamics of a semiflexible filament, which is embedded in a polymer network with a mesh size smaller than the probe filament's contour length. The network is actuated by molecular motors that act as homogeneously dispersed contractile force dipoles. While we approximate the viscoelastic network as a continuum, we assume motor forces to act on the filament at discrete points, where the probe filament is assumed to be coupled mechanically to the meshwork. Without loss of generality, both the meshwork and the filament are described in two dimensions. The filament itself is modeled as an inextensible worm-like chain~\cite{KratkyO.andPorod1949, aragon1985dynamics, chaseReview}. We therefore treat its shape as a continuously differentiable curve in space $\vec{r}(s,t)$, parametrized over time $t$ and its arclength $s$ as sketched in Fig.~\ref{fig:parametrizationOfProbe}.

For every given point in time, we decompose the shape of the probe filament into a transverse and a parallel part with respect to the end-to-end vector $\vec{R}(t)$. This separation yields two internal variables, the transverse $r_\perp(s,t)$ and parallel $r_\parallel(s,t)$ relative position, which are connected by local inextensibility  $\left \lVert \frac{\partial \vec{r}}{\partial s} \right \rVert =1$. The equation of motion governing both variables can be retrieved from a variation of the worm-like chain Hamiltonian $\mathcal{H}=\kappa/2 \int_0^L\mathrm{d}s\, \partial^2 \vec{r}/\partial s^2$, with bending rigidity $\kappa$ and $L$ denoting the total arc length of the filament. We here focus on the linear dynamics of transverse deviations at a given arc length $s$ along the polymer, described by
\begin{align}
  \label{eq:equation of motion}
  \int \limits_{-\infty}^t \,\mathrm{d}t'\,  \alpha(t-t')  r_\perp(t')& = - \kappa \frac{\partial^4 r_\perp}{\partial t^4} +\xi + f_\text{M}.
\end{align}
where the usual stochastic term $\xi(s,t)$ models the net thermal force exerted by the surroundings of the filament consisting of polymer network and solvent. 
The viscoelastic kernel $\alpha$ in Eq.~(\ref{eq:equation of motion}) is related to the bulk shear modulus $G$ via the generalized Stokes theorem~\cite{Gittes1997}
\begin{align} 
  \hat{\alpha}(\omega)&= k_0  \hat{G}(\omega) \label{eq:viscoelasticMemoryShearModulus}.
\end{align}
where $k_0\approx 4 \pi/\ln(L/d) $ is a geometric factor, which also  appears in the transverse drag coefficient $\gamma$ of a infinitesimal rod segment $\gamma \approx k_0\eta$~\cite{Howard2001} in a medium with viscosity~$\eta$. 

The thermal force has zero mean $\langle \xi(s,t)\rangle=0$ and a temporal power spectrum satisfying~\cite{Lau2003}
\begin{align}
\langle \hat{\xi}(s,\omega)\overline{\hat{\xi}}(s',\omega)\rangle= \frac{2k_BT}{\omega} \delta(s-s')\mathrm{Im}\left[\hat{\alpha}(\omega)\right]. \label{eq:noiseSpectrum}
\end{align}

In order to account for the impact of motor forces produced by myosin motors in the vicinity of the probe filament, we include an additional force term $f_\text{M}(s,t)$ in Eq.~(\ref{eq:equation of motion})~\cite{Everaers1999}. In a steady-state scenario, motor protein action gives rise to a fluctuating, but stationary profile of non-thermal forces along the backbone of the probe.
We assume the probe filament to be stationary and to not reptate through the network. This can be achieved, for instance, by linking the probe filament at some points to the network. Reptational movement of the probe filament would cause sampling from a changing spatial motor force profile (see Fig.~\ref{fig:networks}), and thus lead to a blurring of the non-equilibrium dynamics we seek to describe. 
The structure of a crosslinked network is characterized by its mesh-size, which determines the density $\ell_\text{M}$ of motor-probe interaction points $s_n$. We model the motor-induced force  as 
\begin{align}
  \label{eq:motorForce}
  f_\text{M}(s,t) = \sum\limits_{n=1}^{N_\text{M}}\, f_n g\left(s-s_n \right)\mathcal{T}_n(t)
\end{align}
with $N_\text{M}$ denoting the number of entanglement points where motors affect the probe filament, $f_n$ denoting the impact of motor $n$ and $g(s)$ a general spatial kernel describing how a motor impacts on the filament. For simplicity, we choose here a point-like spatial kernel, i.e. $g(s)= \delta(s)$, consistent with experiments that suggest a rather narrow force profile~\cite{Brangwynne2008}. The temporal profile of force generated by an individual motor $\mathcal{T}(t)$ in Eq. \eqref{eq:motorForce} is described as a telegraph process.

Non-muscle myosin motors act as oligomeric complexes and produce forces between actin filaments with correlation times of order 10 s~\cite{Fakhri2014, Guo2014}. This time scale is well separated from temporal correlations of thermal noise, but is on the order of the relaxation times of bending modes that play the main role here. To simplify our theoretical description,  we assume motors to instantaneously develop maximal force with characteristic switching rates between the {\em off} and the {\em on} state $1/\tau_\text{on}$ and vice-versa  $1/\tau_{\rm off}$, as proposed in~\cite{PhysRevLett.100.018104}. A telegraph process $\mathcal{T}(t)$ switches between zero and one without memory and can therefore be considered a simple model of a molecular motor.

The activity of different motors is assumed to be uncorrelated, while the autocorrelation of a given motor telegraph process is exponential in lag time, such that we obtain~\cite{Gardiner2009}
\begin{align}
  \label{eq:telegraphCorrelation}
  \langle \mathcal{T}_n(t)\mathcal{T}_m(t') \rangle = C_1+C_2\delta_{n,m} e^{-\frac{|t-t'|}{\tau_\text{M}} }
\end{align}
with dimensionless constants $C_1=\tau_{\rm off}^2/\left(\tau_{\rm on}+\tau_{\rm off}\right)^2$ and $C_2=\tau_{\rm on}\tau_{\rm off}/\left(\tau_{\rm on}+\tau_{\rm off}\right)^2$. In addition, we defined a motor timescale $\tau_\text{M}^{-1}=\tau_\text{on}^{-1}+\tau_\text{off}^{-1}$. A Fourier transformation of Eq.~(\ref{eq:telegraphCorrelation}) yields a Lorentzian power spectrum in accord with literature~\cite{PhysRevLett.100.018104}. In other words, for high frequencies the motor force spectrum follows a power law $S_\mathcal{T}(\omega) \sim \omega^2$, whereas for lower frequencies, the power spectrum becomes essentially white-noise-like $S_\mathcal{T}\sim \text{const}$.


\section{Characterization of non-equilibrium mode dynamics }
\label{sec:char-noneq-mode}
We start to analyze filament mode dynamics by expanding $r_\perp(s,t)$ in orthogonal eigenmodes $y_q(s)$ of the beam operator in Eq.~(\ref{eq:equation of motion}) as $r_\perp(s,t)=L \sum_q a_q(t)y_q(s)$. This expression implies a choice of units: spatial modes $y_q(s)$ and mode amplitudes $a_q(t)$ have dimensions of $1/{\rm length}^{1/2}$ and ${\rm length}^{1/2}$ respectively. Projecting Eq.~(\ref{eq:equation of motion}) onto a particular spatial mode $y_q(s)$ leads to the  equation of motion in mode space
\begin{align}
   \int \limits_{-\infty}^t \,\mathrm{d}t'\, \alpha(t-t')  a_q(t') &= -\kappa q^4 a_q(t)+ \xi_q(t)+f_{\text{M},q}(t), \label{eq:modeEquationOfMotion}
\end{align}
where we made the implicit assumption, that motor and thermal forces do not affect each other.
Indexed quantities denote projected variables, such as the projected thermal noise $\xi_{q}(t)=L^{-1}\int_0^L\,\mathrm{d}s' \xi(s',t)y_q(s')$. We note that thermal forces of different modes do not correlate $\langle \xi_q \xi_w \rangle \propto \delta_{q,w}$ due to the orthogonality of the spatial modes $y_q(s)$. Indices $q$ and $w$ refer to the corresponding wave vectors, which are usually discretized. For instance, for the relevant case of free-end boundary conditions, $q$ can assume values $q(n)\approx (n+\frac{1}{2})\pi/L$ for any integer $n$. 
Furthermore, the projected motor-induced force can be written as
\begin{align}
  f_{\text{M},q}(t) &= \sum\limits_{n=1}^{N_\text{M}} f_n\left( \delta\ast y_q\right)(s_n)\mathcal{T}_n(t) \nonumber \\  
  &=\sum\limits_{n=1}^{N_\text{M}} f_n y_q(s_n)\mathcal{T}_n(t).\label{eq:projectedMotorForce}
\end{align}

Experimentally accessible quantities, such as mode correlations may now be computed with the projected quantities. To this end, we Fourier transform equation (\ref{eq:modeEquationOfMotion}) in time and obtain
\begin{align}
  \label{eq:fourierEOM}
  \hat a_q\left( \omega\right) &=\hat{\chi}_q\left(\omega\right)\left(\hat{ \xi}_q\left(\omega \right )+\hat{ f}_{\text{M},q}\left(\omega\right) \right).
\end{align}
with the mode response function $\hat{\chi}_q\left(\omega\right)~=~\left(  \hat \alpha \left( \omega\right) + \kappa q^4\right)^{-1}$. Although motor-induced forces may surpass thermal forces by orders of magnitude, we choose to keep both terms in Eq.~(\ref{eq:fourierEOM}). This allows us to smoothly transition between purely thermal and purely active dynamics.
In the thermal case, the above equation together with the assumed thermal noise spectrum Eq.~(\ref{eq:noiseSpectrum}) yields the Fluctuation-Dissipation theorem of mode~$q$
  \begin{align}
    \langle |\hat{a}_q(\omega)|^2\rangle^{\text{Th}}  &= \frac{2k_B T}{\omega} \mathrm{Im}\left[ \hat{\chi}_q(\omega)\right]. \label{eq:modePSD}
  \end{align}
As a first application of our model, we characterize the physiologically relevant case of motor-induced fluctuations of relatively stiff filaments in a viscoelastic medium, such as microtubules in actin-myosin networks~\cite{Brangwynne2008, Fakhri2014}. In particular, we calculate the deviation from the equilibrium mode variance shown in Eq.~(\ref{eq:modePSD}).
The viscoelastic response of crosslinked networks of semiflexible filaments typically shows two regimes of the complex shear modulus, a high-frequency regime where $\hat{G}(\omega)\sim \omega^{3/4}$ \cite{PhysRevE.58.R1241} and a low-frequency plateau regime~\cite{MaggsPlateauModuli, Koenderink2006, IsambertDynamicsRheology, MorseViscoelasticity}. The cytoplasm of living cells exhibits an elastic shear modulus with a weak frequency dependence~\cite{Fabry2001, Fakhri2014} up to ~$100$ Hz. The frequency regime we need to consider here is not only determined by the relaxation times of our probe filament for the modes we can resolve, but also by the timescale of motor activity.  
Experiments suggest, that motor-induced fluctuations become negligible for frequencies higher than $100$ Hz~\cite{Mizuno2007, Mizuno2009}. The processivity time of myosin in cells was measured to be about $5$~s~\cite{Fakhri2014}. In \emph{in vitro} model systems, values for $\tau_\text{off}$ and $\tau_\text{on}$ depend on salt and ATP concentrations. 
Given these facts, we focus on the low-frequency regime and assume, for simplicity, a plateau-like shear modulus $\hat{G}(\omega)$ given by 
\begin{align}
  \label{eq:responseFunction}
  \hat{G} = G_0+i\eta \omega
\end{align}
with an elastic modulus $G_0$ and viscosity $\eta$. 
We can now explicitly calculate correlation functions by applying the Wiener-Khinchin theorem to $\langle \hat{a}_q\overline{\hat{a}}_w \rangle$. This yields the mode correlation function, which decomposes into the usual thermal and an additional motor-induced part
\begin{align}
  \label{eq:generalModeCorrelation}
  \langle a_q(t)a_w(t')\rangle= \langle a_q(t)a_w(t')\rangle^{\text{Th}}+\langle a_q(t)a_w(t')\rangle^{\text{M}}
\end{align}
respectively given by
\begin{align}
  \langle a_q(t)a_w(t')\rangle^{\text{Th}} &=  \frac{k_B T \tau_q}{L^2\gamma}\delta_{q,w}e^{-\frac{\left | t-t'\right | }{\tau_q}}  \label{eq:thermalModeCorrelation}\\
  \langle a_q(t)a_w(t')\rangle^{\text{M}} &= \frac{1}{L^2\gamma^2}\mathbf{F}_{q,w}C_2\mathcal{C}_{q,w}\left( t-t'\right). \label{eq:motorModeCorrelation}
\end{align}
 We have introduced here the mode relaxation time $\tau_q= 1/(\kappa/\gamma q^4+ G_0/\eta)$ and a coupling matrix $\mathbf{F}_{q,w}$.  The elements in this coupling matrix encode the spatial structure of motor-induced forces in terms of bending modes,
\begin{align}
  \label{eq:couplingCoefficient}
\mathbf{F}_{q,w}&= \sum\limits_{n=1}^{N_\text{M}}\, f_n^2 y_q(s_n)y_w(s_n).
\end{align}
In contrast to the purely thermal case, Eq.~(\ref{eq:motorModeCorrelation}) shows that in the active case, modes are no longer independent, but correlate with a magnitude determined by the coupling matrix $\mathbf{F}_{q,w}$ and the function $\mathcal{C}_{q,w}(\Delta t)$ that describes the temporal evolution of the motor-induced mode correlation function.

\begin{align}
 \langle a_q(t)\rangle \langle a_w(t)\rangle&=  \frac{1}{L^2\gamma^2}\tau_q\tau_w C_1 \sum\limits_{n,m}^{N_\text{M}}\, f_nf_m y_q(s_n)y_w(s_m). \label{eq:meanCorrelatorPart}
\end{align}
For simplicity, we assume the average active forces and torques exerted on the filament to vanish, requiring $\langle f_{\rm M}(s,t) \rangle = 0$ for all $s$. This leads to $\sum_nf_n y_q(s_n) = 0$ and thus $\langle a_q(t)\rangle \langle a_w(t)\rangle = 0$.
    
\begin{figure}
 \centering
 \includegraphics[width=1\columnwidth]{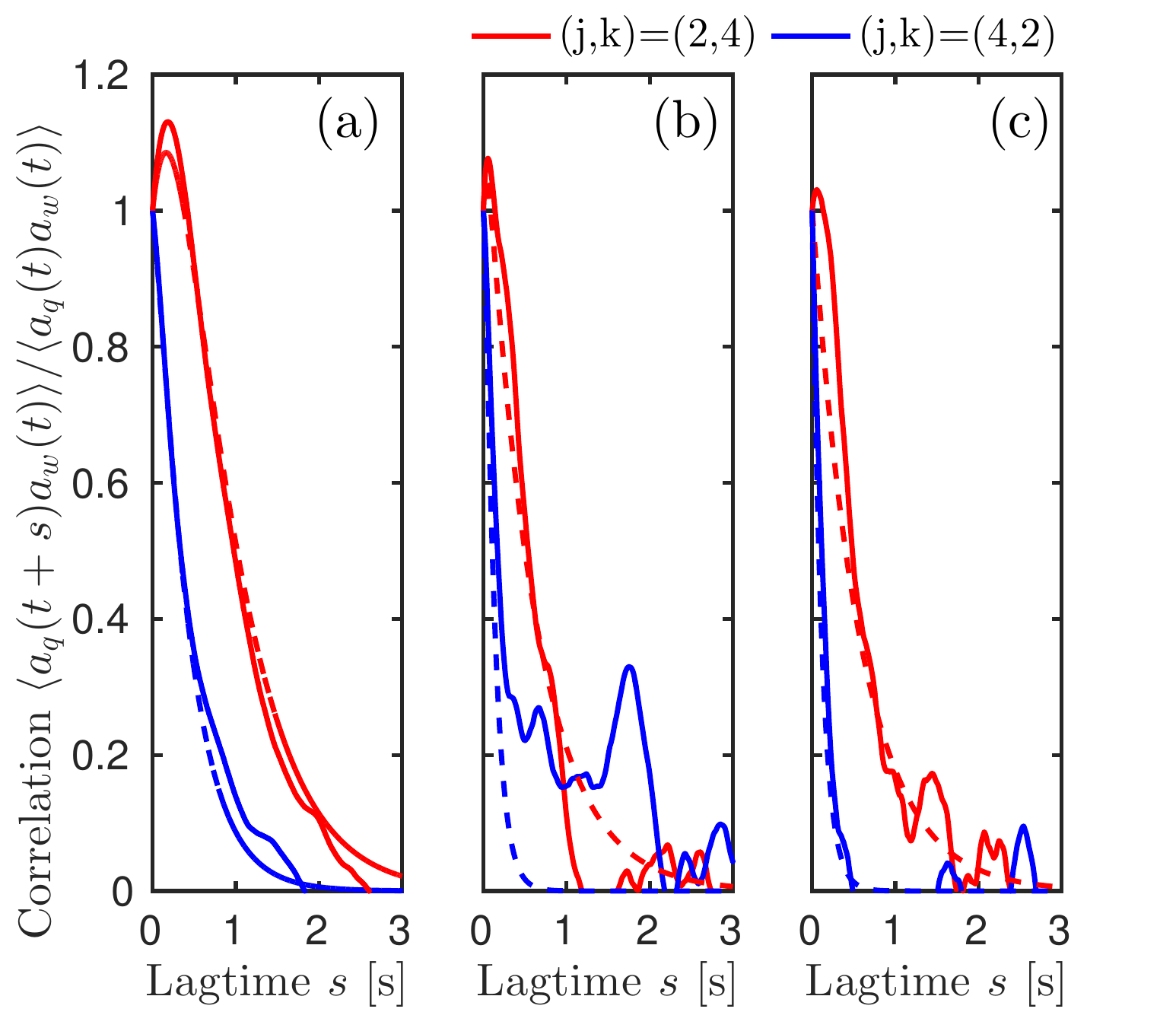}
\caption{The breaking of Onsager's time reversal symmetry results in different non-zero normalized cross-correlations of the mode pair $(j,k)=(2,4)$ (red) and $(j,k)=(4,2)$ (blue). Continuous lines represent correlations inferred from Brownian dynamics simulations, while analytical predictions from Eq.~(\ref{eq:modeMotorCrossCorrelator}) are dashed. The time scale of the motor noise $f_\text{M}(s,t)$ was set to (a) $\tau_\text{M}=0.375$ s, (b) $\tau_\text{M}=0.0375$ s and (c) $\tau_\text{M}=0.000375$ s.}
\label{fig:brokenOnsager}
\end{figure}

The function $\mathcal{C}_{q,w} (t-t')$ in Eq.~(\ref{eq:motorModeCorrelation}) is limited at short times by the motor decorrelation time, is independent of the precise spatial arrangement of the motor-induced forces, and is given by
\begin{align}
    \mathcal{C}_{q,w}( t-t')&= \tau_q\tau_w\left(\frac{e^{-\frac{\lvert  t-t' \rvert }{\tau_\text{M}}}}{\left(1-\frac{\tau_q}{\tau_\text{M}}\right)\left(1+\frac{\tau_w}{\tau_\text{M}}\right)} \right.\nonumber \\
&\left. -2 \frac{\tau_q}{\tau_\text{M}}\frac{e^{-\frac{\lvert t'-t\rvert }{\tau_q}}}{\left(1-\left(\frac{\tau_q}{\tau_{\rm M}}\right)^2\right)\left(1+\frac{\tau_w}{\tau_q}\right)}\right)\label{eq:modeMotorCrossCorrelator}
\end{align}
where $\tau_q$ must be the relaxation time associated with time coordinate $t$, where $t>t'$, and $\tau_w$ is associated with the mode evaluated at~$t'$. While for zero lag time, i.e. $t=t'$, the correlator is symmetric, indices can not be exchanged for $t\neq t'$. Thus, Onsager's time reversal symmetry is broken in this system, as demonstrated in Fig.~\ref{fig:brokenOnsager}~\cite{Onsager1931, Onsager1931a}. The simplest behavior of $\mathcal{C}_{q,w} (t-t')$ can be observed for modes with relaxational time scales $\tau_q$, $\tau_w$ that are much longer than the motor time scale $\tau_\text{M}$. In this limit, $\mathcal{C}_{q,w} (t-t')$ evolves as $\text{Exp}(-\Delta t/\tau_q)$. In Fig.~\ref{fig:brokenOnsager} this transition of the cross-correlation to simple exponential behavior is exemplified for the mode pair $(q,w)=(2, 4)$ for three different motor time scales. By contrast, in equilibrium, these cross-correlations vanish for all lag times $\Delta t$. It is interesting to note that one could therefore directly infer non-equilibrium dynamics from asymmetries in the mode correlation functions. The breaking of the Onsager time reversal symmetry of the mode correlation function is equivalent with the breaking of detailed balance due to motor activity in our model.

As discussed previously~\cite{Gladrow2016}, in the diagonal case, $\mathcal{C}_{q,q}(t-t')$ can be written in a more compact form,
\begin{align}
      \mathcal{C}_{q,q}(t-t')& = \frac{1}{\tau_q^{-2}-\tau_\text{M}^{-2}}\left(e^{-\frac{\left |  t-t'\right|}{\tau_\text{M}}}-\frac{\tau_q}{\tau_\text{M}}e^{-\frac{\left|t-t'\right|}{\tau_q}}\right). \label{eq:modeMotorCorrelator}
\end{align}
We now turn to the coupling matrix $\mathbf{F}_{q,w}$ defined in Eq.~(\ref{eq:couplingCoefficient}). Our motivation for defining these coefficients in this particular way can best be illustrated by regarding the active force cross-correlations $\langle f_q(t)f_w(t')\rangle$ using Eqs.~(\ref{eq:telegraphCorrelation}) and (\ref{eq:motorForce}). Under the assumption that all motor-force processes have the same time scale and interact with the probe in only one location each, we find 
\begin{align}
  \langle f_{\text{M},q}(t)f_{\text{M},w}(t')\rangle &= \frac{1}{L^2} \sum_{i,j} f_i f_j y_q(s_i)y_w(s_j) \langle \mathcal{T}_i(t)\mathcal{T}_j(t') \rangle \nonumber \\
  &= \frac{1}{ L^2}\sum_{i,j} f_i f_j y_q(s_i)y_w(s_j) \delta_{i,j} C_2 e^{-\frac{\lvert t-t'\rvert }{\tau_\text{M}}} \nonumber \\ 
  &= \frac{C_2 e^{-\frac{\lvert t-t'\rvert }{\tau_\text{M}}}}{ L^2}\sum_i f_i^2 y_q(s_i)y_w(s_i) \nonumber \\
  &= \frac{C_2 e^{-\frac{\lvert t-t'\rvert }{\tau_\text{M}}}}{ L^2} \mathbf{F}_{q,w} \nonumber.
\end{align}
Thus the elements of the coupling matrix $\mathbf{F}_{q,w}$ appear in the above equation as the only index-dependent quantities. 

\begin{figure}
  \centering
  \includegraphics[width=1\columnwidth]{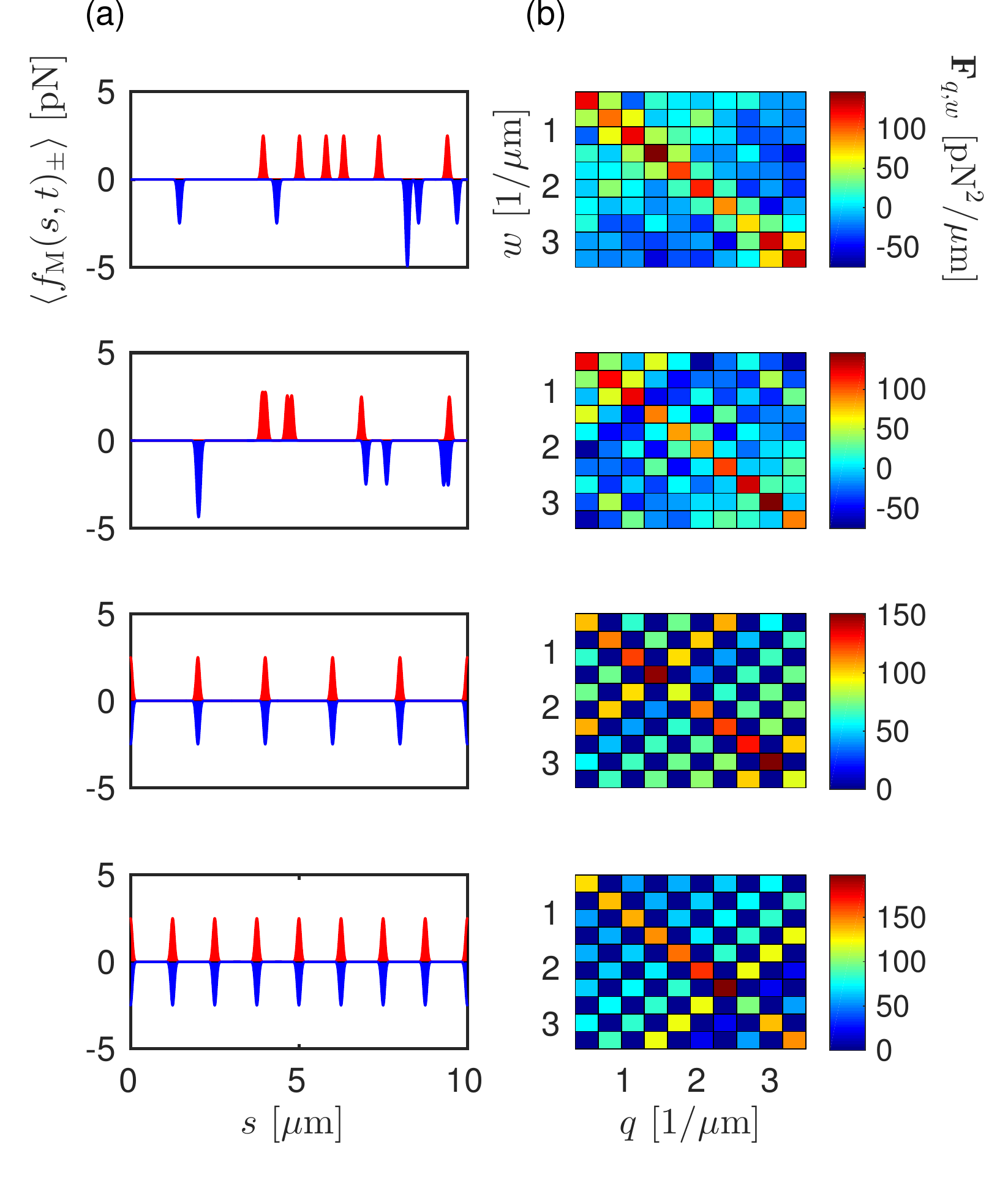}
  \caption{ (color online) (a) Temporal average of upward (blue) and downward (red) motor-induced force $\langle f_\text{M}(s, t) \rangle$. The peaks are shown with a finite width for illustration purposes. In the two top rows, 12 interactions points $s_n$ were chosen from an equal distribution with density $\rho(s_n)=1/L $, half of them were assigned negative force coefficients $f_{-}=-f_{+}$. The magnitude of the force was fixed at $f_{+}=|f_{-}|=10$ pN, $L$ was set to $10 \,\mu\text{m}$ and $\tau_\text{off}/\tau_\text{on}=1/3$. In the two bottom rows, the interaction points were distributed periodically with varying density, such that the average total force and torque were zero. (b) The corresponding coupling matrices $\mathbf{F}_{q,w} $ for the first 11 modes. }
  \label{fig:couplingCoefficientsModeTable}
\end{figure}

The coupling matrix for a given fluctuating filament will depend on the spatial distribution of points through which forces are transmitted to the filament. Fig.~\ref{fig:couplingCoefficientsModeTable} shows three examples of sets of interaction-points $\{s_n\}$ with the associated coupling matrices. The first two are disordered distributions, while the third and fourth example constitute periodic patterns. Since the coupling matrix depends on both the choice of motor-probe interaction strengths $f_n$ and the placement of interaction points $s_n$, one could consider the ensemble average over all possible disordered network-probe interactions~$\langle \mathbf{F}_{q,w}\rangle_\text{dis}$. However, under the assumption of a uniform distribution of possible interaction points $\rho(s_n) = 1/L$, we find $\langle \mathbf{F}_{q,w}\rangle_\text{dis}\propto \delta_{q,w}$ due to the orthogonality of the dynamic normal modes. Consequently, the mode correlations will disappear in this ensemble average in spite of non-equilibrium driving. This result holds true, even when the distances between the interaction points $\Delta s_n = s_{n+1}-s_n$ are drawn from an exponential distribution with some characteristic spatial frequency $1/\ell_{\rm M}$. The ensemble average of the coupling matrix $\langle \mathbf{F}_{q,w}\rangle_\text{dis}$ over disordered network configurations with such an exponential distribution of motor interaction points is still diagonal, despite the structured probe-mesh interaction. This is because the relative position of the probe to the network is a uniformly distributed variable and hence the average force geometry is uniform.

In any case, diagonality of $\mathbf{F}_{q,w}$ implies that both broken Onsager's time reversible symmetry and breaking of detailed balance (see below) will not be visible in normal-mode phase space. In other words, ensemble averaging over trajectories recorded from different probe filaments and thus over different force profiles will tend to conceal the non-equilibrium nature of the dynamics. The same will be true in the limit of high spatial densities of interaction points $\ell_{\rm M} \to 0$, where the sum in Eq.~(\ref{eq:couplingCoefficient}) approaches an integral over $s$,

\begin{align}
  \label{eq:highDensityLimit}
  \mathbf{F}_{q,w}\overset{\ell_{\rm M} \to 0}{\to} \frac{1}{\ell_{\rm M}}\int \limits_{0}^L\mathrm{d}s \, f^2(s)y_q(s)y_w(s).
\end{align}
If force amplitudes don’t spatially vary, the coupling matrix $\mathbf{F}_{q,w}$ therefore also becomes diagonal in the high-density limit~\cite{Gladrow2016}.

\section{Brownian dynamics simulations}
\label{sec:simulations}

To numerically test our analytical predictions, we carried out Brownian dynamics simulation of Eq.~(\ref{eq:equation of motion}). To this end, we approximated the probe filament as a discrete worm-like chain composed of $N$ beads, connected by stiff elastic springs to ensure local inextensibility.  We discretized the fourth-derivative in equation (\ref{eq:equation of motion}) with a central-stencil scheme. Free-end boundary conditions were implemented by endowing the chain with four ghost-beads, two at each end respectively, to which the fourth-derivative stencil was applied. At each step, we positioned the ghost beads along the tangent at the respective end of the filament. 
The elastic confinement of the probe filament was simulated by springs attached to every bead of the chain and to a fixed point at a unit distance extending orthogonally to the local filament tangent. Motor-interaction points $s_n$ were positioned at an equal distance $\ell_\text{M}$ along the filament to apply the assumptions made in the previous section. Motor action was modeled by a telegraph process with a variance as described by Eq.~(\ref{eq:telegraphCorrelation}). In order to obtain mode amplitude traces $\vec{a}(t)$ from the simulated dynamics, we projected the filament backbone onto the free-rod eigenmodes of the beam equation $y_q(s)$~\cite{Brangwynne2007} at each time point $t_i$.  

\section{Active mode dynamics violate detailed balance}
\label{sec:break-deta-balance}
Non-equilibrium driving of a network by embedded motors 
results in an enhancement of fluctuations of an inserted probe filament. We showed above that dynamic normal modes, which are uncorrelated in equilibrium, may start to correlate in such an active environment (Eqs.~(\ref{eq:thermalModeCorrelation}) and (\ref{eq:motorModeCorrelation})). Cross-correlations alone, however, are not a reliable tool to diagnose non-equilibrium or to infer force profiles. Especially lower mode correlations converge rather slowly and tensile effects, which we neglected in our treatment, may result in non-zero cross-correlations in experiments, even in equilibrium.

A more robust measure of non-equilibrium is the breaking of detailed balance. We analyze the breaking of detailed balance by looking for circulating probability currents in normal mode phase space. In particular, we discuss how the structure of such currents depends on the geometry of interactions between our probe filament and the motor-activated network. 

\begin{figure}
  \centering
  \includegraphics[width=0.49\textwidth]{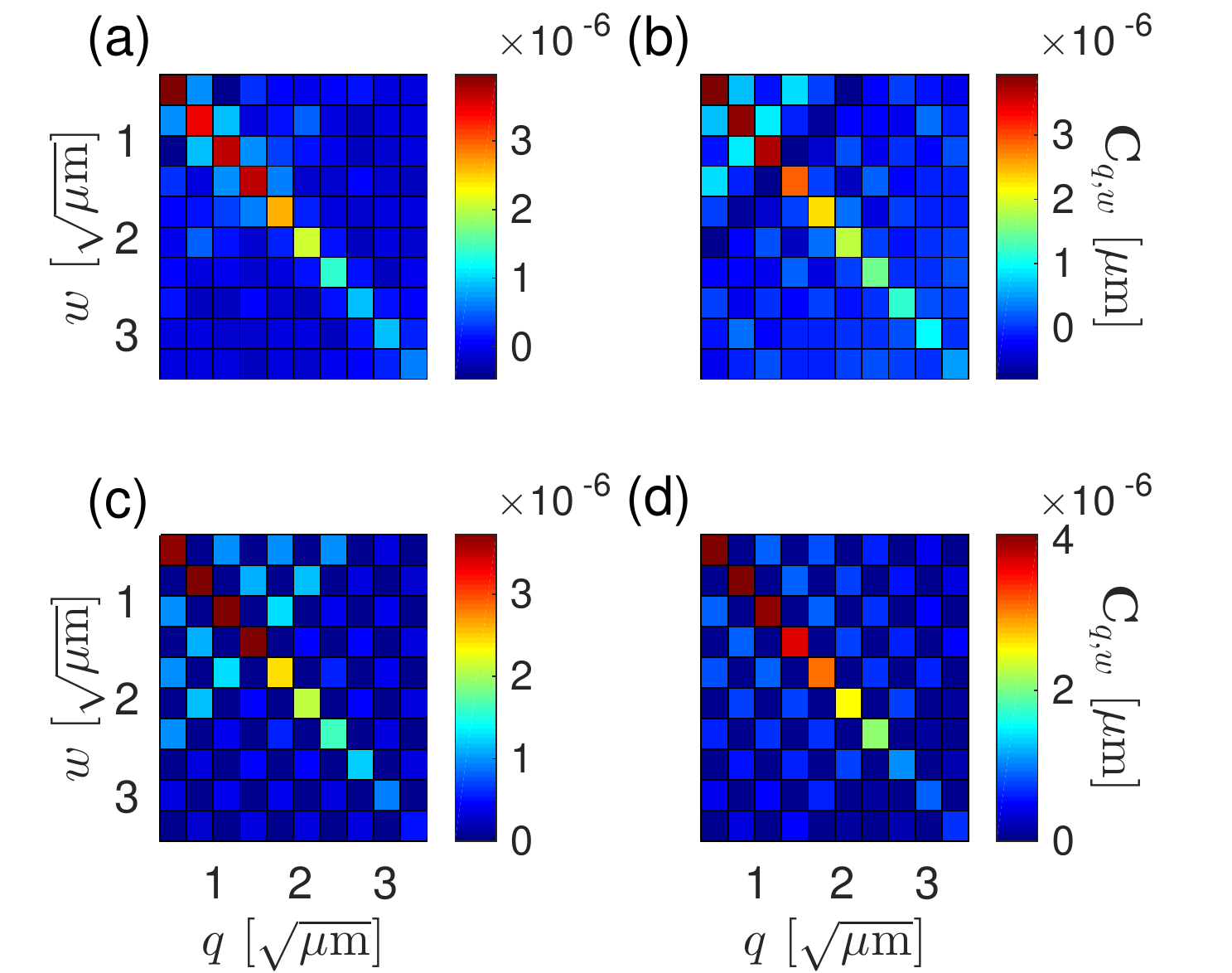}
  \caption{Correlation matrices calculated from Eq.~(\ref{eq:correlationMatrixWhiteNoise}) for the same sets of random interaction points $\{s_n\}$ as in Fig.~\ref{fig:couplingCoefficientsModeTable}.}
  \label{fig:randomCorrelationMatrix}
\end{figure}

While thermal forces drive all modes equally and independently (Eq.~(\ref{eq:thermalModeCorrelation})), the network-probe interactions $f_\text{M}(s,t)$, which are assumed to be distributed heterogenously, excite modes unevenly and in a correlated fashion. In conjunction with different mode relaxation times, this gives rise to a directed probability current in the dynamic normal mode space of the probe filament. We can understand this phenomenon quantitatively in the white-noise limit of the mode equation of motion (see Eq.~(\ref{eq:modeEquationOfMotion})).
\begin{figure}
  \centering
 \includegraphics[width=1\columnwidth]{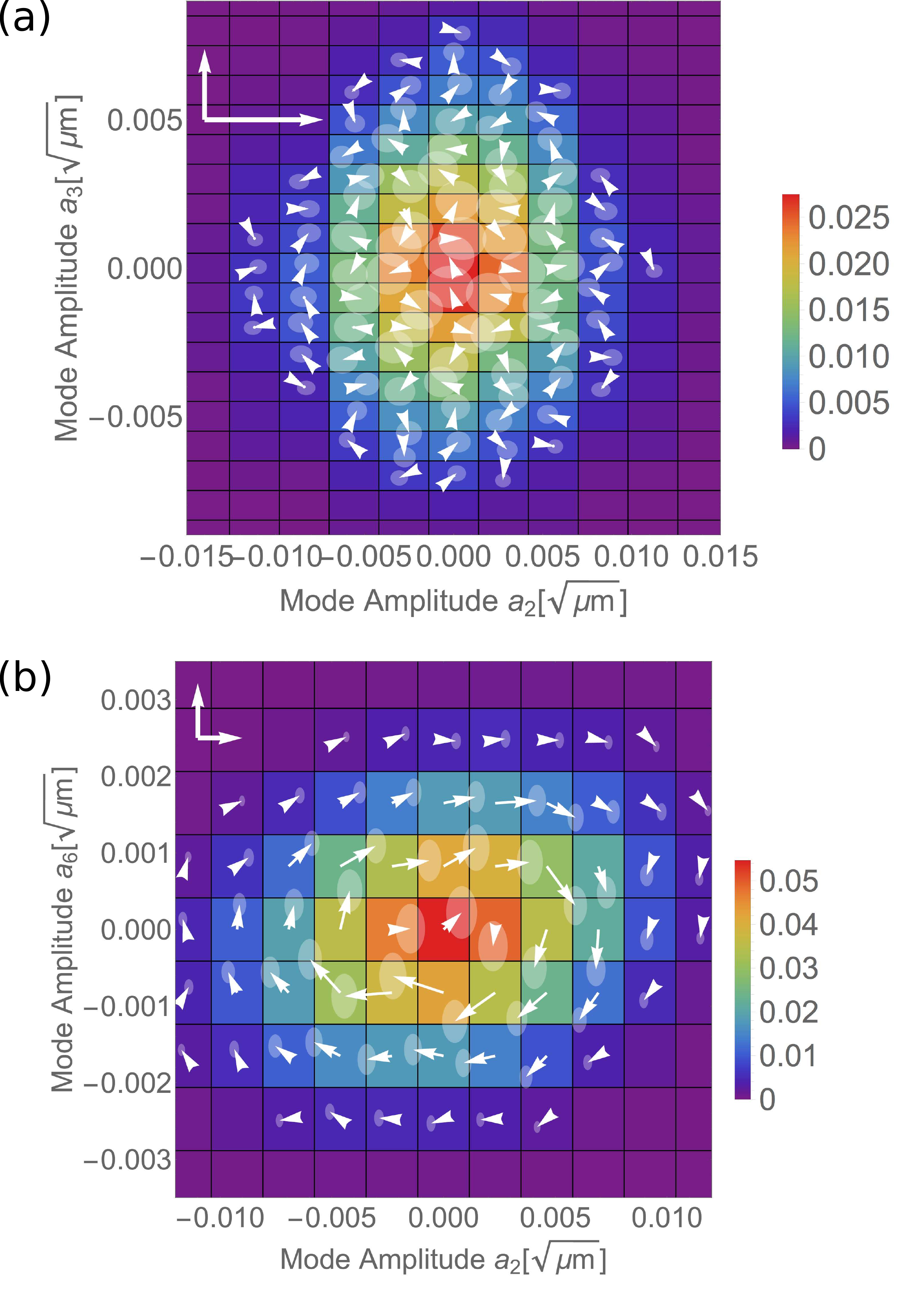}
\caption{Probability fluxes in normal mode phase space inferred from simulations. In both cases the arrow-scale is $0.2 \, 1/\text{s}$. (a) Phase space of mode pair $a_2$ and $a_3$ with no significant probability currents and (b) of mode pair $a_2$ and $a_6$, with significant currents in accord with our theory. The indices $2$, $3$ and $6$ describe the free-end mode numbers and correspond to wave vectors of $q(2)\approx 0.73 \, \mu m^{-1}$, $q(3)\approx 1.1 \, \mu m^{-1}$ and $q(6)\approx 2.04 \, \mu m^{-1} $ respectively. Interaction points $s_n$ were chosen in a periodic pattern as in the third row from the top in Fig.~\ref{fig:couplingCoefficientsModeTable}. Motor parameters are chosen to be the same as in Fig.~\ref{fig:couplingCoefficientsModeTable}, with the exception of $\tau_\text{off}=0.0005$s, $\tau_\text{on}=0.0015$s. Filament parameters were set to model microtubules in a typical cytoskeletal actin-gel: $\kappa=24\times 10^{-24}$ Nm$^2$, $L=10 \, \mu$m, $G_0=10$ Pa and $\eta=1 \,$ Pas.}
  \label{fig:probabilityCurrents}
\end{figure}
On long timescales $T\gg \tau_\text{M}$, the internal motor time scale becomes negligible, and we can approximate the motor forces to appear as white-noise processes. In this limit, we rewrite the correlator of the telegraph process $\langle \mathcal{T}_n(t)\mathcal{T}_m(t') \rangle \to C_1+ \frac{C_2}{2} \delta_{n,m}\tau_\text{M} \delta(t-t')$. The motor time scale $\tau_\text{M}$ here remains only as a proportionality constant of the variance of the active forces to ensure appropriate dimensions.
We can now absorb the motor processes $f_{\text{M},q}(t)$ into the thermal force $\xi_q(t)$, which yields a new white noise process $\psi_q(t)$ with a correlator
\begin{align}
   \langle \psi_q(t)\psi_w(t')\rangle =&\left(4k_BT\gamma\delta_{q,w} +  C_2\tau_\text{M} \mathbf{F}_{q,w} \right) \frac{\delta \left( t-t'\right)}{2L^2} \label{eq:newThermalNoise}.
\end{align}
Furthermore, we again adopt the Kelvin-Voigt description of embedding network and solvent $\hat G(\omega)=G_0+i\omega \eta$, since we now operate in a regime of even lower frequencies. Such a complex shear modulus with constant real part corresponds to a continuum of springs and simple drag penalizing transverse deviations. We can therefore split up the left-hand side of Eq.~(\ref{eq:equation of motion}) into an elastic and viscous force. The Langevin equation of motion for the normal modes Eq.~(\ref{eq:modeEquationOfMotion}) then becomes 
\begin{align}
  \gamma \frac{\mathrm{d} a_q}{\mathrm{d}t}(t)&= -\left( \kappa q^4+ \gamma G_0/\eta\right) a_q(t) + \psi_q(t). \label{eq:whiteNoiseLangevin}
\end{align}
The corresponding diffusion matrix is
\begin{align}
  \label{eq:1}
  \mathbf{D}_{q,w}&= \frac{1}{2\gamma^2}\int\limits_{-\infty}^{\infty} \mathrm{d}t\, \langle \psi_q(t)\psi_w(0)\rangle.
\end{align}
with this result, we are in a position to write down the Fokker-Planck equation of the system
\begin{align}
  \frac{\partial \rho}{\partial t}\left( \vec{a},t\right)&= -\vec{\nabla}\cdot\left [\mathbf{K}\vec{a}\rho\left(\vec{a},t\right)-\mathbf{D}\vec{\nabla}\rho\left(\vec{a},t\right) \right] \label{eq:fpe}
\end{align}
where $\mathbf{K}_{q,w}=-1/\tau_q \delta_{q,w}$ denotes the deterministic matrix. The steady-state solution of the above equation is a Gaussian probability density $\rho(\vec{a})= \mathcal{N}^{-1}e^{-\frac{1}{2}\vec{a}^T\mathbf{C}^{-1}\mathbf{a}} $ with a normalization constant~$\mathcal{N}^{-1}$ and a correlation matrix defined by
\begin{align}
 \label{eq:explicitCorrelationMatrix}
 \mathbf{C}_{q,w}=\langle a_q(t)a_w(t) \rangle^{\psi },
\end{align}
where the superscript indicates that the average is taken in the white-noise limit over $\psi(t)$ defined in Eq.~(\ref{eq:newThermalNoise}).
We can analytically obtain the white-noise correlation matrix $\mathbf{C}$ either from Eq.~(\ref{eq:fpe}) or directly from a power expansion of Eq.~(\ref{eq:motorModeCorrelation}) up to linear order in $\tau_\text{M}$
\begin{align}
  \label{eq:correlationMatrixWhiteNoise}
  \mathbf{C}_{q,w}&= \frac{1}{\gamma^2L^2}\left(2k_BT\gamma \delta_{q,w} +C_2\tau_\text{M} \mathbf{F}_{q,w} \frac{\tau_q\tau_w}{\tau_q+\tau_w}\right).
\end{align}
For a random collection of motor interaction points $\{s_n\}$, drawn from a constant probability density over the filament length $L$, the coupling matrix and thus the correlation matrix may exhibit off-diagonal elements as exemplified in Fig.~\ref{fig:randomCorrelationMatrix}.

Since the correlation matrix $\mathbf{C}_{q,w}$ in Eq.~(\ref{eq:correlationMatrixWhiteNoise}) is proportional to the coupling matrix $\mathbf{F}_{q,w}$, the relative values of the entries of this matrix can, in principle, be inferred from the correlation, if mode relaxation times $\tau_j$ are known. A direct comparison of Fig.~\ref{fig:couplingCoefficientsModeTable} and Fig.~\ref{fig:randomCorrelationMatrix} reveals how the motor force distribution $f_\text{M}(s,t)$ translates into mode cross-correlations.

The quantity in brackets in Eq.~(\ref{eq:fpe}) can be identified as the probability current $\vec{j}(\vec{a})$, which in the steady-state limit becomes
\begin{align}
 \vec{j}(\vec{a}) = (\mathbf{K}+\mathbf{D}\mathbf{C}^{-1})\vec{a}\rho(\vec{a}) \equiv \mathbf{\Omega} \vec{a}\rho(\vec{a}), \label{eq:current}
\end{align}
with $\mathbf{\Omega}$ denoting a matrix of frequencies, which we discuss below. By inserting the result for $\mathbf{C}$ into Eq.~(\ref{eq:correlationMatrixWhiteNoise}), we can directly obtain the current $\vec{j}(\vec{a})$.
\begin{figure}
 \centering
 \includegraphics[width=0.49\textwidth]{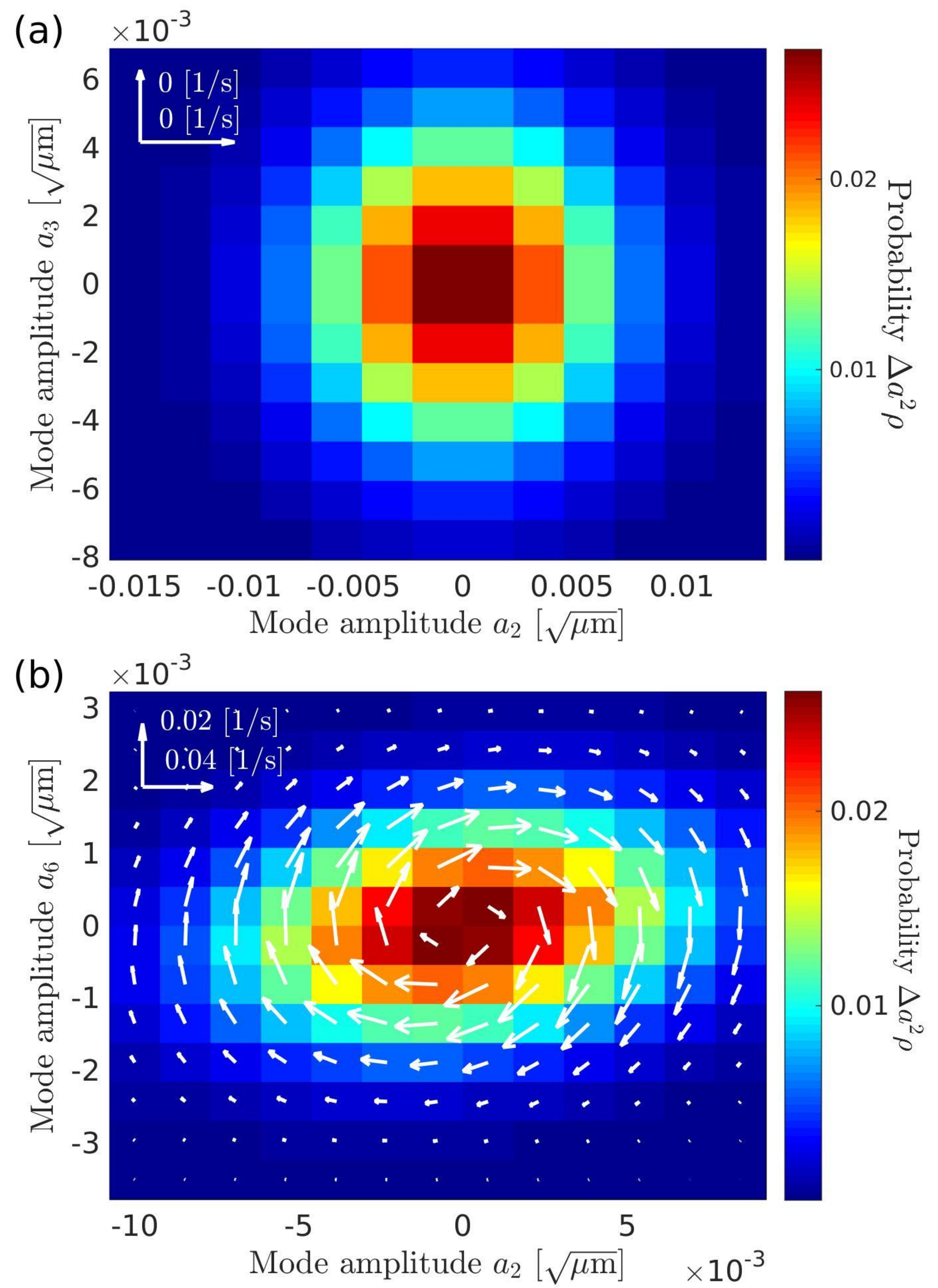}
   \caption{Analytical predictions of the currents for (a) mode pair $a_2$ and $a_3$ and (b) mode pair $a_2$ and $a_6$. All parameters are chosen to match the scenario presented in Fig.~\ref{fig:probabilityCurrents}. }
  \label{fig:combinedTheoryCurrents}
\end{figure}
The probability currents arising from regularly spaced interaction points $s_n$ are plotted in Fig.~\ref{fig:combinedTheoryCurrents}.
We calculated these currents in two-dimensional reduced mode spaces $a_q\times a_w$, where dynamics of other modes were disregarded for simplicity. Due to the symmetry of the motor induced forces $f_\text{M}(s)$ in Fig.~\ref{fig:probabilityCurrents} and Fig.~\ref{fig:combinedTheoryCurrents}, currents develop only in pairs of mode numbers of the same parity, for instance odd-odd, as one would expect from the structure of the coupling matrix (see third panel in Fig.~\ref{fig:couplingCoefficientsModeTable}).

In our simulations, we measured cumulative probability currents $\vec{J}_{q,w} $ over small rectangles $\square_{n,m}$ in the corresponding $a_q \times a_w$ subspace. More precisely, we applied the current estimator proposed in~\cite{Battle2016} to all simulated two-mode amplitude trajectories $\{a_q(t_i), a_w(t_i)\}_i$. In Fig.~\ref{fig:probabilityCurrents}, coarse-grained probability currents inferred from Brownian dynamics simulations are shown. Following the bootstrapping technique described previously~\cite{Battle2016}, we calculated error estimates (white ellipses) in order to distinguish significant currents.

\section{Cycling Frequencies in the White-Noise Approximation}
\label{sec:break-deta-balance}

If a system violates detailed balance it does so in any coordinate system; it should not depend on the choice of coordinates whether it is in equilibrium or not.  In fact, detailed balance requires the product $\mathbf{K}\mathbf{D}$ to be symmetric~\cite{Weiss2003}, which is not fulfilled for general $\mathbf{F}_{q,w}$.
In a steady-state, the current $\vec{j}(\vec{a})$ must necessarily be purely rotational to fulfill $\vec{\nabla}\cdot\vec{j}(\vec{a}) =0$. A rotational probability current manifests itself in a preferred angular direction of the underlying stochastic dynamics of modes $\vec{a}(t)$. Trajectories sampled from such a state must therefore exhibit, on average, a cycling motion. 
The characteristic frequencies $\omega_{q,w}$ of this cycling motion are, in principle, experimentally accessible and can be considered a scalar measure of irreversible dynamics.

In a two-dimensional space, the stochastic dynamics of the system on average cycles at a single characteristic frequency, which can be calculated analytically as the imaginary part of the first of the two nonzero eigenvalues of the frequency matrix $\mathbf{\Omega}$~\cite{Gladrow2016, Weiss2003, Ghanta2017}.
In higher-dimensional systems, however, the situation is less clear. Without prior knowledge about the structure of the motor-generated forces, we can transform the observed mode traces $\vec{a}(t)$ into so-called {\em correlation-identity coordinates} $\widetilde{a}_q(t)$,  where $\langle \widetilde{a}_q(t) \widetilde{a}_w(t) \rangle = \widetilde{\mathbf{C}}_{q,w}=\delta_{q,w}$. This is achieved by multiplying the mode vector $\vec{a}(t)$ by the matrix square-root of the inverse-correlation matrix $\sqrt{\mathbf{C}^{-1}}$. We note that the wave-vector indices $q, w,...$ become dimensionless in correlation-identity coordinates. 

In this coordinate system, theoretically meaningful cycling frequencies can be estimated by projecting the system onto two-dimensional hyperplanes, disecting mode space. If we followed only $\widetilde{a}_q(t)$ and $\tilde{a}_w(t)$, we would perceive an apparent frequency $\omega_{q,w}$, which, in general, would  be different from any eigenvalue of the frequency matrix~$\tilde{\mathbf{\Omega}}$, but can nevertheless be calculated. Indeed, as shown below, the apparent cycling frequency in this hyper-plane spanned by correlation-identity coordinates is given by the matrix elements of $\tilde{\mathbf{\Omega}}_{q,w}$, which here read~$(\widetilde{\mathbf{K}}+\widetilde{\mathbf{D}})_{q,w}$.

\begin{figure}
 \centering
 \includegraphics[width=1\columnwidth]{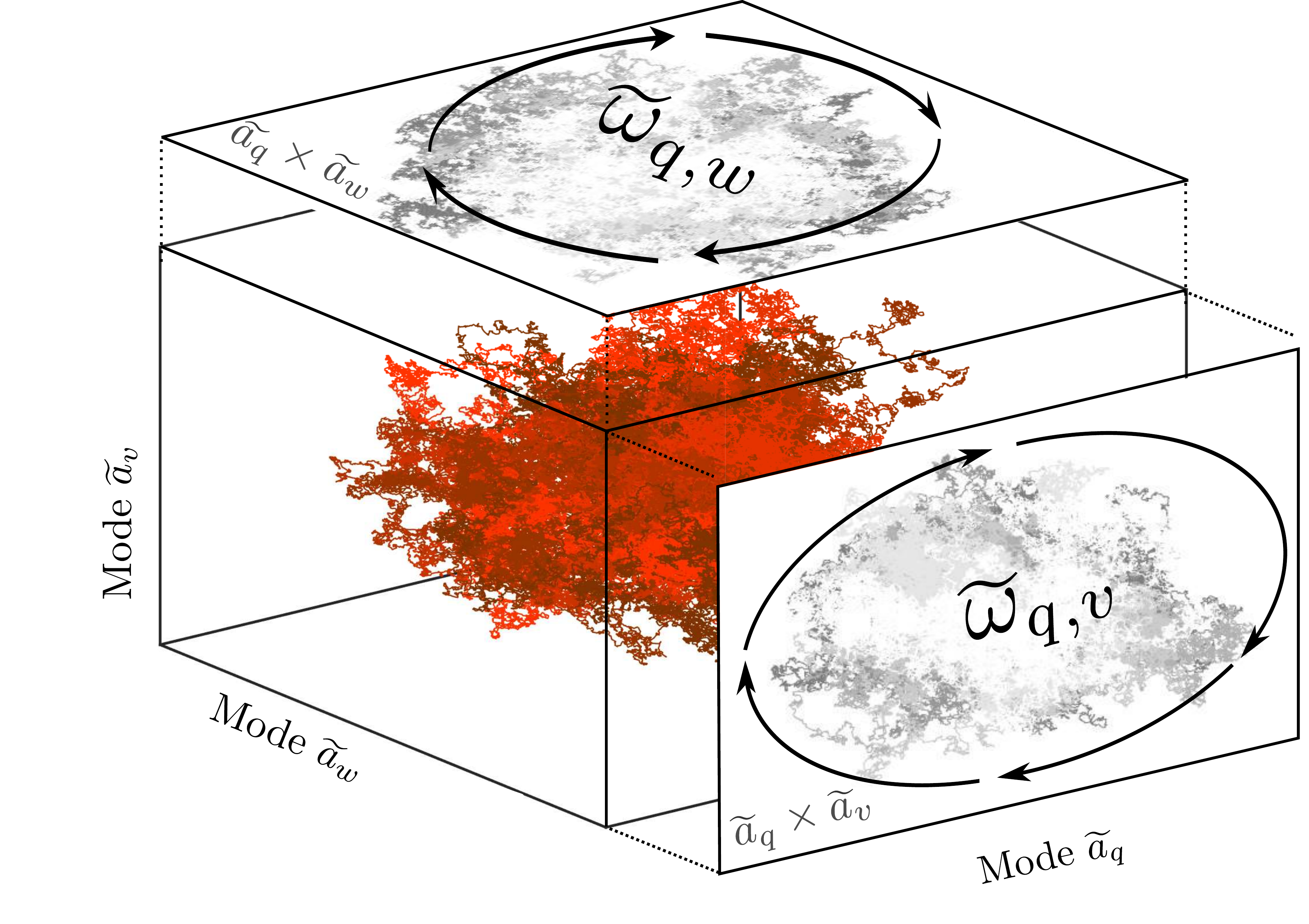}
 \caption{The apparent cycling frequency depends on the choice of the observed plane. An observer of the plane $\widetilde{a}_q \times \widetilde{a}_v$ will measure the frequency $\widetilde{\omega}_{q,v}=\widetilde{\mathbf{\Omega}}_{q,v}$, while $\widetilde{\omega}_{q,w}=\widetilde{\mathbf{\Omega}}_{q,w}$ would be measured in the plane $\widetilde{a}_q \times \widetilde{a}_w$.} 
\end{figure}

We begin by showing that integrating out unobserved variables in correlation-identity coordinates effectively reduces the matrix $\tilde{\mathbf{\Omega}}$ to those elements $\tilde{\mathbf{\Omega}}_{q,w}$ corresponding to the two observed degrees of freedom. We indicate $\widetilde{a}_q\times \widetilde{a}_w$ hyperplane quantities by a superscript $(2)$. 
\begin{align}
\widetilde{j}^{(2)}_q (\widetilde{a}_q,\widetilde{a}_w)&=\int\limits_{-\infty}^\infty \, \mathrm{d}\widetilde{a}_{q_1} \dots \int\limits_{-\infty}^\infty \, \mathrm{d}\widetilde{a}_{q_N} \widetilde{j}_q(\vec{\widetilde{a}}) \nonumber\\
&=\int\limits_{-\infty}^\infty \, \mathrm{d}\widetilde{a}_{q_1} \dots \int\limits_{-\infty}^\infty \, \mathrm{d}\widetilde{a}_{q_N} \sum_v \widetilde{\mathbf{\Omega}}_{q,v} \widetilde{a}_v \widetilde{\rho}(\vec{\widetilde{a}}) \nonumber \\
&=\widetilde{\mathbf{\Omega}}_{q,w}\widetilde{a}_w \, (2\pi)^{-1}e^{-\frac{\widetilde{a}_q^2+\widetilde{a}_w^2}{2}} \label{eq:reducedDensity}
 \end{align}
where the dots indicate integrals over every degree of freedom, except $\widetilde{a}_q$ and $\widetilde{a}_w$.  Here, we used that $\mathbf{\Omega}$ is skew-symmetric in correlation-identity coordinates, i.e. $\widetilde{\mathbf{\Omega}}_{q,w}=-\widetilde{\mathbf{\Omega}}_{w,q}$~\cite{Weiss2003}. 

Therefore, the current $\widetilde{j}^{(2)}=( \widetilde{j}^{(2)}_q, \widetilde{j}^{(2)}_w)^T$ on the $\widetilde{a}_q\times \widetilde{a}_w$ hyperlane is given by the lower dimensional analogue to Eq.~(\ref{eq:current}), $\widetilde{j}^{(2)}= \widetilde{\mathbf{\Omega}}^{(2)}\vec{\widetilde{a}}\widetilde{\rho}^{(2)}(\vec{\widetilde{a}})$ with $\widetilde{\mathbf{\Omega}}^{(2)}$ being an antisymmetric $2\times 2$ matrix with off-diagonal elements $\widetilde{\mathbf{\Omega}}_{q,w}$ and $-\widetilde{\mathbf{\Omega}}_{q,w}$. The term $(2\pi)^{-1}e^{-\frac{\widetilde{a}_q^2+\widetilde{a}_w^2}{2}}$ in Eq.~(\ref{eq:reducedDensity}) is simply the marginal joint distribution $\widetilde{\rho}^{(2)}$ of $\widetilde{a}_q$ and $\widetilde{a}_w$. The eigenvalues of $\widetilde{\mathbf{\Omega}}$ hence read $\lambda_{\pm} = \pm i \widetilde{\mathbf{\Omega}}_{q,w}$, which implies that $\widetilde{j}^{(2)}(\widetilde{a}_q,\widetilde{a}_w)$ is purely rotational. From this analysis, we conclude that the apparent cycling frequency in the $\widetilde{a}_q \times \widetilde{a}_w$ hyperplane in correlation-identity coordinates is given by
\begin{align}
 \widetilde{\omega}_{q,w} & = \widetilde{\mathbf{\Omega}}_{q,w}. \label{eq:apparentCyclingFreq}
\end{align}
We note that the matrix elements of $\widetilde{\mathbf{\Omega}}_{q,w}$ must be calculated in the full-dimensional system.

\begin{figure}[h]
  \centering
  \includegraphics[width=1\columnwidth]{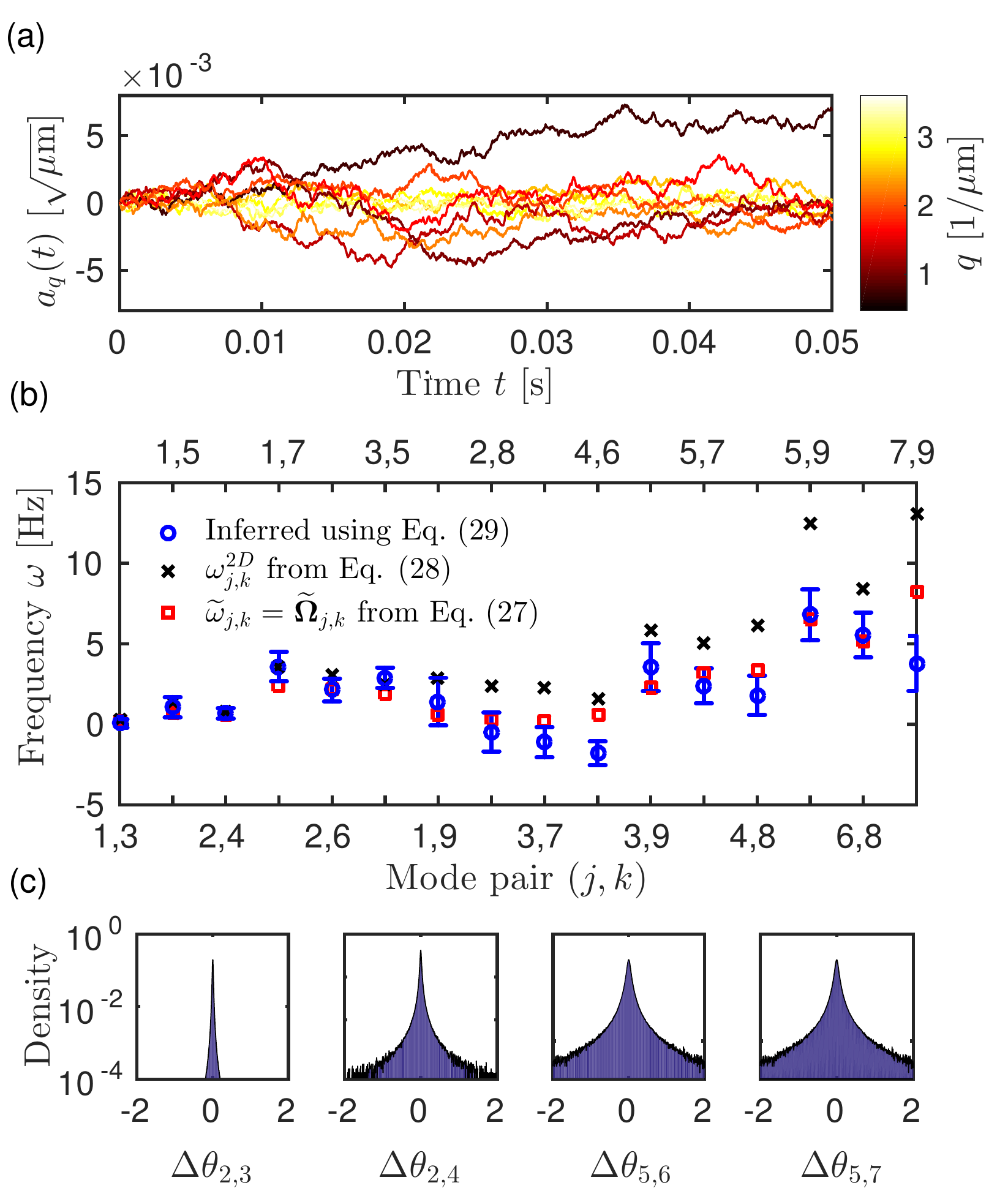}
  \caption{(a) The first eleven mode amplitudes $a_q(t)$ inferred from the first 3000 steps of the simulation. (b) Comparison of net cycling frequencies calculated analytically using Eq.~(\ref{eq:netCyclingFrequency}) (black $x$) for isolated pairs of modes of the same parity and numerically in a 15-dimensional mode space from the matrix elements $\widetilde{\mathbf{\Omega}}_{j,k}$ (red squares) in correlation-identity coordinates as described in Eq.~(\ref{eq:apparentCyclingFreq}). The blue circles are frequencies inferred from the same Brownian simulation data as in Fig.~\ref{fig:probabilityCurrents} using Eq.~(\ref{eq:freqEstimator}). We excluded all data points with a radius $\sqrt{\widetilde{a}_q^2+\widetilde{a}_w^2}<0.1$, where small steps across the origin would lead to large angular changes. Error bars indicate the standard error of the mean. (c) Distribution of angular displacements in log-scale for a few mode pairs.}
  \label{fig:netCyclingFrequencies}
\end{figure}

In mode spaces with more than two dimensions, we obtained frequencies numerically (see Fig.~\ref{fig:netCyclingFrequencies}). To derive analytical predictions for the net cycling frequency, we now consider a two-dimensional system constructed from $a_q \times a_w$. Here, we do not need to transform the system into correlation-identity coordinates because a 2D system only cycles at a single frequency. Intuitively, this simple 2D model should result in an overestimate of the corresponding frequency of the actual higher-dimensional system since this approach excludes transitions out of the considered plane, which would tend to reduce the in-plane cycling frequencies. Indeed, the two-dimensional frequency $\omega_{q,w}^\text{2D}$ appear to be an upper bound for the actual frequencies in the higher-dimensional system, as shown in Fig.~\ref{fig:netCyclingFrequencies}.

We may, nevertheless, use the approximative calculation to explore under which circumstances we expect detailed balance to be broken.
In our two-dimensional subsystem, $\omega_{q,w}^\text{2D}$ can be directly calculated as the positive eigenvalue $\lambda_+$ of $\mathbf{\Omega}$ in Eq.~(\ref{eq:current}) computed for a system consisting only of modes $a_q$ and $a_w$ and reads
\begin{widetext}
\begin{align}
  \omega_{q,w}^\text{2D} &=  \frac{  \left (\tau_q -\tau_w\right)\mathbf{F}_{q,w} }{\sqrt{\tau_q\tau_w\left( \left(\tau_q+\tau_w\right)^2 \left( \left(\frac{2k_B T\gamma}{C_2\tau_M}\right)^2+\frac{2k_B T \gamma}{C_2 \tau_\text{M}}\left(\mathbf{F}_{q,q}+\mathbf{F}_{w,w} \right)+\mathbf{F}_{w,w}\mathbf{F}_{q,q}  \right) -4\tau_q\tau_w \mathbf{F}_{q,w}^2\right) }}.\label{eq:netCyclingFrequency}
\end{align}  
\end{widetext}
this result shows that for (i) diagonal coupling matrices or (ii) equal relaxation times, detailed balance is always maintained. Furthermore, Eq.~(\ref{eq:netCyclingFrequency}) illustrates how the thermal noise floor, which is constant along the filament, affects the non-equilibrium cycling and may even conceal it in the limit of weak motor action.

As predicted, detailed balance appears to hold, i.e. not be broken, in planes of even and odd modes, for example $a_2 \times a_3$, for the regular geometry of interaction $\{s_n\}$. The magnitude and structure of the currents numerically calculated for the pairs $a_2$, $a_6$ and $a_3$ and $a_5$ match our analytical predictions in Fig.~\ref{fig:probabilityCurrents}. For higher modes, which correspond to shorter length scales, the finite width of the motor-filament interaction in the simulation results in a coupling of even and odd mode amplitudes.

From an experimental point-of-view, where noise in the data is often limiting, it is desirable to determine the value of a single statistical variable instead of spreading sampled data over a number of variables. The net cycling frequency $\omega_{q,w}$ or the cross-correlations $\langle a_q(t)a_w(t')\rangle$ are examples of such single-variable non-equilibrium indicators. The net cycling frequency in a plane $a_q\times a_w$ can be inferred from empirical mode traces using the definition $\omega_{q,w}= \langle \dot{\phi}_{q,w}\rangle$, where $\phi_{q,w}(t)=\text{atan}(a_w(t)/a_q(t))$ is the polar angle. This leads to the observed frequency in the plane  
\begin{align}
  \omega_{q,w}&= \langle \frac{ \dot{a}_w(t) a_q(t) -  \dot{a}_q(t) a_w(t)}{a^2_q(t) +a^2_w(t)}\rangle \label{eq:freqEstimator}.
\end{align}

In Fig.~\ref{fig:netCyclingFrequencies} (a), we compare the prediction for the two-dimensional net cycling frequencies $\omega^\text{2D}$ based on Eq.~(\ref{eq:netCyclingFrequency}) with the prediction for the apparent frequencies in a high-dimensional space from Eq.~(\ref{eq:apparentCyclingFreq}) and frequencies obtained from simulations. The distributions of angular displacements $\Delta \theta$ are shown on a log-scale in Fig.~\ref{fig:netCyclingFrequencies} (b). As the distributions widen with increasing mode number, it becomes more difficult to precisely estimate the average net cycling frequency. 

While it is tempting to interpret $\omega_{q,w}$ as a metric for the distance to equilibrium, the cycling frequencies are actually determined by two features of the system: the natural relaxation times of the respective modes and the strength of driving relative to the thermal background. Thus, the cycling frequencies provide a comprehensive measure of the non-equilibrium dynamics of different scales.

\section{Summary}
\label{sec:summary}
In this paper, we extended existing models describing the stochastic dynamics of semiflexible probe filaments embedded in viscoelastic media to incorporate effects of non-equilibrium force generation in active media. This model would, for example, describe probe 
filaments such as microtubules or single-walled carbon nanotubes in the cell's actin cytoskeleton activated by myosin motors.
In particular, we derived analytical descriptions of dynamic normal mode cross- and autocorrelations that can be used to characterize the stochastic forces the probe filaments are subject to in the active networks under steady state conditions.

We showed that motor-induced forces may lead to probability currents in phase spaces spanned by dynamic normal mode amplitudes, indicating a breaking of detailed balance. The structure of these currents is closely related to the active force profile along the backbone of the probe. We confirmed these analytical results with currents inferred from Brownian dynamics simulations of semi-flexible filaments driven by molecular motors. In steady-state, we showed that these divergence-free currents form a net cyclic probability flux in the phase space of mode amplitudes with a characteristic frequency $\omega_{q,w}$. We discussed implications of the dimensionality of mode space on the relation between measured frequencies and active force profiles and derived an approximate analytical expression for these cycling frequencies in the white-noise limit of motor action. Furthermore, we indicated special cases in which we expect detailed balance to hold in spite of non-equilibrium driving: In planes $a_q\times a_w$ spanned by modes with (i) equal relaxation times or (ii) vanishing correlations, detailed balance will hold.

Many predictions of our model, such as the magnitude of probability currents, depend on the geometry of the network around the probe filament, which might be difficult to chart. By contrast, variables like relaxation times or correlation functions of the normal mode amplitudes appear to be more robust against geometrical details of the active forcing.
We therefore introduced the breaking of Onsager's time reversal symmetry as a novel non-equilibrium marker, which only involves temporal aspects of the correlation function. In fact, we show that under non-equilibrium conditions, reciprocity of the correlation function breaks down. This can be used as a powerful tool to detect and quantify non-equilibrium behavior using the dynamics of probe filaments.

In summary, mode cross-correlations and cycling frequencies, may serve in the future to detect non-equilibrium dynamics and characterize the spatial distribution and the temporal behavior of motor-induced forces acting on individual probe filaments. Estimating probability currents from sample trajectories requires rather large data sets, however. The availability of photostable fluorescent filaments, such as single-walled carbon nanotubes, will bring the proposed experiments within practical reach. 
More generally, our results could be applied to any extended object with non-thermal fluctuations such as chromosomes, membranes, cellular organelles, whole cells or even tissues.

\begin{acknowledgments}
We thank F. MacKintosh, N. Fakhri, F. Mura, F. Gnesotto, and G. Gradziuk for helpful discussion. This research was supported by the National Science Foundation under Grant No. NSF PHY11-25915,by the German Excellence Initiative via the program NanoSystems Initiative Munich (NIM) (C.P.B.), the GRK2062 (C.P.B) and the Deutsche Forschungsgemeinschaft (DFG) Collaborative Research Center SFB 937 (Project A2), the European Research Council Advanced Grant PF7 ERC-2013-AdG, Project 340528 (C.F.S), and the Cluster of Excellence and DFG Research Center Nanoscale Microscopy and Molecular Physiology of the Brain (CNMPB) (C.F.S.). 
\end{acknowledgments}
\bibliographystyle{apsrev4-1} 
\end{document}